\documentstyle[osa,twocolumn]{revtex}
\begin{document}
\title{Critical behavior and the Neel temperature of quantum quasi-two-dimensional
Heisenberg antiferromagnets}
\author{V.Yu.Irkhin$^{*}$ and A.A.Katanin}
\address{620219, Institute of Metal Physics, Ekaterinburg, Russia.}
\maketitle

\begin{abstract}
The nonlinear sigma-model and its generalization on $N$-component spins, the
$O(N)$ model, are considered to describe thermodynamics of a quantum
quasi-two-dimensional (quasi-2D) Heisenberg antiferromagnet. A comparison
with standard spin-wave approaches is performed. The sublattice
magnetization, Neel temperature and spin correlation function are calculated
to first order of the $1/N$-expansion. A description of crossover from a
2D-like to 3D regime of sublattice magnetization temperature dependence is
obtained. The values of the critical exponents derived are $\beta =0.36,$ $%
\eta =0.09$. An account of the corrections to the standard logarithmic term
of the spin-wave theory modifies considerably the value of the Neel
temperature. The thermodynamic quantities calculated are universal functions
of the renormalized interlayer coupling parameter. The renormalization of
interlayer coupling parameter turns out to be considerably temperature
dependent. A good agreement with experimental data on La$_2$CuO$_4$ is
obtained. The application of the approach used to the case of a ferromagnet
is discussed.
\end{abstract}


\section{Introduction}

Last time, great interest is paid to properties of quasi-two-dimensional
(quasi-2D) antiferromagnets in connection with the investigations of layered
perovskites \cite{Joungh} and copper-oxide systems, including high-$T_c$
superconductors. In particular, La$_2$CuO$_4$ gives one of the best known
example of a quasi-2D system with small magnetic anisotropy. Unlike 2D
systems, quasi-2D ones have finite values of magnetic ordering temperature.
At small interlayer couplings $J^{\prime }$ the value of magnetic transition
temperature is small in comparison with the intraplane exchange parameter $J$%
. There are a number of approximations which enable to describe the
thermodynamics of such systems. The standard spin-wave theory (SWT) takes
into account only the spin-wave excitations which exist for quasi-2D systems
in a wide temperature range up to about $J$ (Refs.\cite{Yoshioka,Takahashi}%
). SWT does not take into account the dynamic and kinematic interaction
between spin waves, which are important at temperatures near magnetic phase
transition point. By this reason, SWT gives too high values of magnetic
transition temperature. Recently, the self-consistent spin-wave theory \cite
{Yoshioka,Takahashi,ArovasBook} (SSWT) has been proposed which takes into
account partially the interaction between spin waves. However, the value of
the Neel temperature in SSWT is still too high in comparison with
experiment, and the critical behavior is described quite incorrectly.

To describe the magnetic phase transition we have to take into consideration
fluctuation (non-spin-wave) corrections to thermodynamic quantities. It is
difficult to take into account such corrections in the standard technique of
the Green's functions because of essentialy nonlinear character of equations
of motion. There exists the interpolation approximation by Tyablikov \cite
{Tyablikov} which is based on the random-phase decoupling of equations of
motion for the transverse spin Green's function$.$ This approach yields
often results which are roughly satisfactory from the experimental point of
view. At the same time, it is difficult to justify and improve such
approximations.

To develop perturbation theory which describes correctly the critical
behavior, we have to introduce a formal large parameter in the Heisenberg
model. Thus the Heisenberg model can be treated as a model with a large
degeneracy within the $1/N$-expansion. This expansion may be introduced in
two different ways. The first way \cite{Read,Starykh,Starykh1,OurCP} treats
the Heisenberg model as a particular case ($M=2$) of the $SU(M)$ model (i.e.
of the model with $M$ states per spin degree of freedom at each site). Since
the $M\rightarrow \infty $ limit corresponds to SSWT (see, e.g., Ref.\cite
{ArovasBook}), at finite $M$ thermodynamics is described in terms of the
spin-wave picture of excitation spectrum. The second way \cite
{Brezin,Chubukov} is to consider the Heisenberg model as a particular case ($%
N=3$) of the $O(N)$ model (i.e. of the model with $N$-component spins). The
limit $N\rightarrow \infty $ gives the quantum spherical model and the large-%
$N$ case corresponds to the fluctuation (non-spin-wave) picture. The
advantage of the $1/N$ (or $1/M$) expansions over, say, the quasiclassical $%
1/S$ expansion is their applicability near the phase transition temperature.

Since $N=3$ and $M=2$ are in fact not large, the convergence of such
expansions must be investigated separately. For low-dimensional magnets with
$d=2$ (see Ref. \cite{Starykh}) and $d=2+\varepsilon $ (Ref. \cite{OurCP})
the results in the $SU(M)$ model coincide in the zeroth order in $1/M$ with
those of the one-loop RG analysis, and in the first order in $1/M$ with the
results of the two-loop RG analysis. In these cases the $1/M$-corrections to
thermodynamic quantities are small. However, quasi-2D systems belong to $3D$
symmetry group so that corresponding $1/M$-corrections are not small (see
discussion in Ref. \cite{OurCP}) and the series in $1/M$ is poorly
convergent. Unlike the $1/M$-expansion in the $SU(M)$ model, the first-order
$1/N$-corrections in the $O(N)$ model, which were considered in the quantum
2D case \cite{Chubukov} and in the classical case at an arbitrary
dimensionality $2<d<4$ (see, e.g., Ref. \cite{Hikami}), lead to results
which are close to those obtained by other methods. The applicability of the
$1/N$-expansion at arbitrary dimensionality $2\leq d\leq 4$ is important for
the investigation of quasi-2D systems since they demonstrate the dimensional
crossover from 2D to 3D behavior (see, e.g., Ref.\cite{Joungh}). On the
other hand, the renormalization group $\varepsilon $-expansion is not
appliciable for $d=2$ and $d=3$ simultaneously: for $\varepsilon =d-2$ it
cannot describe satisfactorily the case $d=3$ and vice versa, for $%
\varepsilon =4-d$ the behavior at $d\rightarrow 2$ is poor.

Thus, instead of direct calculation of corrections to SSWT, we start in this
paper from the quantum spherical model, $O(\infty )$ and then find the $1/N$%
-corrections. Although the results in the $O(\infty )$ and $SU(\infty )$
models are different, it will be shown that already in the first order in $%
1/N$ at low enough temperatures the results in the $O(N)$ model are
identical to those in $SU(\infty )$ (i.e. in SSWT). At higher temperatures
the results of SSWT are modified due to fluctuation corrections.

The plan of the paper is as follows. In Sect.2 we review various
approximations in the theory of quasi-2D systems, which are based on the
spin-wave picture of excitation spectrum, and analyze the corresponding
expressions for the Neel temperature. In Sect.3 we formulate the $O(N)$
model for the quasi-2D case and the technique of the $1/N$-expansion, which
is a generalization of that by Chubukov et al \cite{Chubukov} for the 2D
case. In Sect.4 we calculate the magnetization, Neel temerature and spin
correlation function to first order in $1/N.$ In Sect.5 we discuss our
results and compare them with experimental data on La$_2$CuO$_4$.

\section{Spin-wave approximations in the theory of quasi-2D antiferromagnets}

We start from the Heisenberg Hamiltonian of a quasi-2D antiferromagnet
\begin{equation}
H=\frac 12\sum_{ij}J_{ij}{\bf S}_i{\bf S}_j  \label{Heis}
\end{equation}
with the exchange interactions $J_{i,i+\delta }=J$ for $\delta $ in a plane
and $J_{i,i+\delta }=J^{\prime }$ for $\delta $ perpendicular to the planes.

At small values of interlayer coupling $J^{\prime }$ it is possible to
derive analytical results for the Neel temperature. First we consider the
standard spin-wave theory. The spectrum of spin waves has the form
\begin{equation}
E_{{\bf q}}^{\text{SWT}}=S(J_0^2-J_{{\bf q}}^2)^{1/2}  \label{SpSw}
\end{equation}
where $J_{{\bf q}}$ is the Fourier transforms of the exchange parameter
\begin{equation}
J_{{\bf q}}=2J(\cos q_x+\cos q_y)+2J^{\prime }\cos q_z
\end{equation}
The sublattice magnetization is determined by
\begin{equation}
\overline{S}=S+\frac 12-\sum_{{\bf q}}\frac{J_0S}{2E_{{\bf q}}}\coth \frac{%
E_{{\bf q}}}{2T}
\end{equation}
For small values of $J^{\prime }/J$ SWT yields different analytical
expressions for the Neel temperatue in the quantum regime ($T_{\text{Neel}%
}\ll JS$) and classical regime ($T_{\text{Neel}}\gg JS$). We have
\begin{eqnarray}
T_{\text{Neel}}^{\text{SWT}} &=&4\pi JS^2\times   \nonumber \\
&&\ \ \left\{
\begin{array}{cc}
1/\ln (T_{\text{Neel}}^2/8JJ^{\prime }S^2) & \ln (J/J^{\prime })\gg 2\pi S
\\
1/\ln (Jq_0^2/J^{\prime }) & 1\ll \ln (J/J^{\prime })\ll 2\pi S
\end{array}
\right.   \label{TNSW}
\end{eqnarray}
Here $q_0\simeq \pi $ is a cutoff parameter determined by the boundary of
the Brillouin zone. Note that for the quantum case the main contribution to
integrals over the wavevector comes from the region with $q\leq T\,$, while
in the classical case the value of $T_{\text{Neel}}$ is determined by the
whole Brillouin zone.

The spin-wave spectrum in SSWT and the Tyablikov approach is renormalized in
different ways. SSWT \cite{Yoshioka,Takahashi,ArovasBook} takes into account
the interaction between spin waves in the simplest self-consistent Born
approximation. There exist several generalizations of SSWT on quasi-2D
systems \cite{Our1st,OurFMM,Liu}. We will follow to approach of Refs.\cite
{Our1st,Liu} which gives more satisfactory results at small $J^{\prime }/J$.
The spin-wave spectrum in SSWT has the form
\begin{eqnarray}
E_{{\bf q}}^{\text{SSWT}} &=&S\left( \gamma _0^2-\gamma _{{\bf q}}^2\right)
^{1/2} \\
\gamma _{{\bf q}} &=&2\gamma (\cos q_x+\cos q_y)+2\gamma ^{\prime }\cos q_z
\nonumber
\end{eqnarray}
where $\gamma $ and $\gamma ^{\prime }$ are the renormalized exchange
parameters which are determined from the self-consistent equations
\begin{eqnarray}
\gamma /J &=&\sum_{{\bf q}}\frac{\gamma _{{\bf q}}S}{E_{{\bf q}}}\cos
q_x\coth \frac{E_{{\bf q}}}{2T}+2\overline{S}(T)  \label{g} \\
\gamma ^{\prime }/J^{\prime } &=&\sum_{{\bf q}}\frac{\gamma _{{\bf q}}S}{E_{%
{\bf q}}}\cos q_z\coth \frac{E_{{\bf q}}}{2T}+2\overline{S}(T)  \label{g'}
\end{eqnarray}
The sublattice magnetization is given by
\begin{equation}
\overline{S}=S+\frac 12-\sum_{{\bf q}}\frac{\gamma _0S}{2E_{{\bf q}}}\coth
\frac{E_{{\bf q}}}{2T}
\end{equation}
At small values of $J^{\prime }/J$ we have
\begin{equation}
\overline{S}=\overline{S}_0-\frac T{4\pi \gamma S}\times \left\{
\begin{array}{cc}
\ln (T^2\,/8\gamma \gamma ^{\prime }S^2) & S(JJ^{\prime })^{1/2}\ll T\ll JS
\\
\ln (\gamma q_0^2/\gamma ^{\prime }) & JS\ll T\ll JS^2
\end{array}
\right.  \label{MagnSSWT}
\end{equation}
where $\overline{S}_0$ is the sublattice magnetization in the ground state.
The quantity $\gamma $ varies slowly with temperature in the whole region $%
T<T_{\text{Neel}}$ and may be replaced by its zero-temperature value $\gamma
(0).$ According to Refs.\cite{Yoshioka,Takahashi}, we have
\begin{equation}
\gamma (0)=1.1571J,\,\,\,\,\overline{S}_0=0.3034  \label{gs}
\end{equation}
for $S=1/2$ and $\gamma (0)=J,$ $\overline{S}_0=S$ $\,$for $S\rightarrow
\infty $. The second case in (\ref{MagnSSWT}) may be realized only in the
classical limit $S\gg 1.$ One can see from (\ref{MagnSSWT}) that the value
of the critical exponent for the magnetization is $\beta _{\text{SW}}=1.$
The same critical behavior takes place at an arbitrary $d>2.$ This result is
correct only at $d=2+\varepsilon $ to leading order in $\varepsilon $ ($%
\beta =1-2\varepsilon ,$ see, e.g., Refs.\cite{Brezin,OurCP}), and for
higher dimensionalities $\beta <1$.

As follows from (\ref{MagnSSWT}), the Neel temperature is determined by the
equation of Ref. \cite{Our1st} (note that some coefficients in this paper
are incorrect)
\begin{eqnarray}
T_{\text{Neel}}^{\text{SSWT}} &=&4\pi \gamma _cS\overline{S}_0
\label{TNSSWT} \\
&&\times \left\{
\begin{array}{cc}
1/\ln (T_{\text{Neel}}^2\,/8\gamma _c\gamma _c^{\prime }S^2) & \ln
(J/J^{\prime })\gg 2\pi S \\
1/\ln (q_0^2\gamma _c/\gamma _c^{\prime }) & 1\ll \ln (J/J^{\prime })\ll
2\pi S
\end{array}
\right.   \nonumber
\end{eqnarray}
Here $\gamma _c\simeq \gamma (0)$ and $\gamma _c^{\prime }$ are the
renormalized exchange parameters at $T=T_{\text{Neel}}$, The value of $%
\gamma _c^{\prime }$ determined from (\ref{g'}) is
\begin{equation}
\gamma _c^{\prime }=(T_{\text{Neel}}/4\pi \gamma _cS^2)J^{\prime }
\label{grSSWT}
\end{equation}
in both the quantum and classical regimes. Note that the renormalization of
the interlayer coupling in (\ref{TNSSWT}) plays a crucial role in lowering
the Neel temperature in comparison with its SWT value (\ref{TNSW}) since $%
\gamma _c\gamma _c^{\prime }/JJ^{\prime }=T_{\text{Neel}}/4\pi JS^2\ll 1$.

In the Tyablikov theory \cite{Tyablikov} (TT) the excitation spectrum has
the form
\begin{equation}
E_{{\bf q}}^{\text{TT}}=\overline{S}(J_0^2-J_{{\bf q}}^2)^{1/2}.
\label{SpTyab}
\end{equation}
As well as in a ferromagnet, the proportionality of the spectrum to $%
\overline{S}$ is not quite correct at low temperatures: in the
antiferromagnet the spin-wave frequency varies as $T^4$, while the
sublattice magnetization as $T^2$ (see, e.g., Ref.\cite{SWT}). The equation
for $\overline{S}$ at $S=1/2$ reads
\begin{equation}
1/\overline{S}=\sum_{{\bf q}}\frac{J_0\overline{S}}{E_{{\bf q}}}\tanh \frac{%
E_{{\bf q}}}{2T}
\end{equation}
and has a more complicated form for higher spins \cite{Tyablikov}. Near the
Neel temperature TT yields at arbitrary $S$ and any space dimensionality $%
d>2 $%
\begin{equation}
\overline{S}=\left[ \frac{2\Gamma _ST_{\text{Neel}}^{\text{TT}}}{SJ_0}\left(
1-\frac T{T_{\text{Neel}}^{\text{TT}}}\right) \right] ^{1/2}  \label{MagnTbT}
\end{equation}
where $\Gamma _S$ is some function of $S$, $\Gamma _{1/2}=3$. Thus, unlike
SSWT, the critical exponent for the magnetization has the standard
mean-field value, $\beta _{\text{TT}}=1/2$. For small $J^{\prime }/J,$ TT
yields
\begin{equation}
T_{\text{Neel}}^{\text{TT}}\simeq \frac{4\pi JS^2}{\ln (Jq_0^2/J^{\prime })}
\label{TNTYAB}
\end{equation}
The result (\ref{TNTYAB}) is lower than SSWT value (\ref{TNSSWT}) and closed
to experimental data (see Sect.5). On the other hand, the result (\ref
{TNTYAB}) coincides with that of the spherical model (which is adequate only
in the classical limit $S\rightarrow \infty $ \cite{Baxter,Nagaev}) and with
the result of the spin-wave approximation (\ref{TNSW}) in the classical
regime $T_{\text{Neel}}\gg JS.$ The Tyablikov approximation gives the same
result (\ref{TNTYAB}) (with the replacement $J\rightarrow -J,\,J^{\prime
}\rightarrow -J^{\prime }$) for the Curie temperature of a ferromagnet ($%
J,J^{\prime }<0$)$.$ This demonstrates that near the critical temperature TT
does not take into account quantum fluctuations which are important for
small values of $S$. Thus we may conclude that TT is satisfactory from the
practical, but not from theoretical point of view.

To leading logarithmic accuracy, all the discussed approaches give the same
value of the Neel temperature. However, this accuracy is insufficient to
treat experimental data. In particular, the factor of $q_0^2\sim 10$ in the
classical regime is often not taken into account (see, e.g., Ref.\cite
{Joungh}), although this factor gives an essential contribution to $T_{\text{%
Neel}}$.

To improve the description of the critical region and obtain a better
appoximation for the Neel temperature in the quantum case, it is necessary
to take into account fluctuation corrections to the spin-wave theory result
for $T_{\text{Neel}}$ (\ref{TNSW}) more correctly than in SSWT and TT. To
this end we use in the next Sections the $1/N$-expansion in the $O(N)$ model.

\section{The quantum nonlinear sigma-model and $O(N)$ model for quasi-2D
quantum antiferromagnets}

To describe thermodynamics of quantum antiferromagnets we consider the
nonlinear sigma-model which was proposed for the one-dimensional Heisenberg
model in Ref.\cite{Haldane}. In the 2D case this model was applied in Refs.
\cite{Chakraverty,Chubukov}. The large value of the correlation length $\xi
\gg a$ ($a$ is the lattice parameter in the plane) plays the crucial role in
the Haldane's mapping of an antiferromagnetic Heisenberg model (\ref{Heis})
to quantum nonlinear sigma-model (see e.g. Ref. \cite{ArovasBook}). This
gives a possibility to separate and integrate out the ``fast'' modes with
space scale $l\leq \Lambda ^{-1}$ ($\Lambda $ satisfies to $\xi ^{-1}\ll
\Lambda \ll a^{-1}$) retaining ``slow'' modes with $l>\Lambda ^{-1}$.

In the quasi-2D case we have $\xi (T\leq T_{\text{Neel}})=\infty .$ However,
at small $q$ we have
\begin{equation}
J_0-J_{{\bf q}}\simeq J(aq)^2+2J^{\prime }(1-\cos q_z)
\end{equation}
Thus besides the ``true'' correlation length $\xi ,$ there exists also
another variable with scaling dimensionality of length
\begin{equation}
\xi _{J^{\prime }}=1/\alpha ^{1/2}\gg a  \label{ksi}
\end{equation}
where $\alpha =2J^{\prime }/Ja^2$ is the interlayer coupling parameter; in
this paper we consider only the case where $\alpha \ll 1$. On the scale of
order of $\xi _{J^{\prime }}$ the regime of fluctuations changes from $2D$
to $3D$ one. Thus we may use the scale $\xi _{J^{\prime }}$ to separate
``fast'' and ``slow'' modes in the Haldane's mapping. Depending on the value
of the imaginary time slab thickness
\begin{equation}
L_\tau =c/T  \label{Lt}
\end{equation}
($c\sim JSa$ is the fully renormalized spin-wave velocity), three regimes
are possible\

\noindent (i) $L_\tau \sim \xi _{J^{\prime }},$ or, equivalently, $T\sim
\alpha ^{1/2}c\sim (JJ^{\prime })^{1/2}.$ This is an analog of the quantum
critical regime $L_\tau \sim \xi $ in the 2D case \cite{Chakraverty,Chubukov}

\noindent (ii) $a\ll L_\tau \ll \xi _{J^{\prime }},$ i.e. $\alpha ^{1/2}c\ll
T\ll c$ which is an analog of the renormalized classical regime $a\ll L_\tau
\ll \xi $ in the 2D case and

\noindent (iii) the classical regime $L_\tau \ll a$ (i.e. $JS\ll T$).

Since the regime (i) is well described by the standard spin-wave theory (or
by SSWT), we do not treat the thermodynamics at temperatures of order of $%
(JJ^{\prime })^{1/2}$. From (\ref{TNSW}) and (\ref{TNSSWT}) one can see that
$T_{\text{Neel}}\gg (JJ^{\prime })^{1/2}.$ In the regimes (ii) and (iii)
implementation of principles of finite-size scaling gives
\[
T_{\text{Neel}}=\rho _s\Phi (\xi _{J^{\prime }}/L_\tau ,\xi _{J^{\prime
}}/a)
\]
were $\rho _s\sim JS^2$ is the fully renormalized spin stiffness, $\Phi
(x,y) $ is a scaling function with $\Phi (\infty ,\infty )=0.$ In the regime
(ii) we have $\xi _{J^{\prime }}/L_\tau \ll \xi _{J^{\prime }}/a,$ so that
\begin{equation}
T_{\text{Neel}}=\rho _s\Phi (\xi _{J^{\prime }}/L_\tau ,\infty )=\rho _s\Phi
_{\text{q}}(T_{\text{Neel}}/\alpha ^{1/2}c)  \label{TNSCAL}
\end{equation}
while in the regime (iii) $\xi _{J^{\prime }}/L_\tau \gg \xi _{J^{\prime
}}/a $ and
\begin{equation}
T_{\text{Neel}}=\rho _s\Phi (\infty ,\xi _{J^{\prime }}/a)=\rho _s\Phi _{%
\text{cl}}(1/\alpha ^{1/2}a)  \label{TNSCAL1}
\end{equation}
Note that the results of SWT\ (\ref{TNSW}) and SSWT (\ref{TNSSWT}) for the
Neel temperature agree with (\ref{TNSCAL}) for the quantum regime and with (%
\ref{TNSCAL1}) for the classical regime. At the same time, the result of the
Tyablikov approximation (\ref{TNTYAB}) satisfies the classical regime
scaling form (\ref{TNSCAL1}) for all spin values, which confirms the absence
of quantum fluctuations at the critical temperature in this approximation.
As follows from (\ref{TNSCAL1}), the value of Neel temperature in the
classical regime depends on fluctuations on a scale of order of lattice
constant, i.e. is non-universal. Therefore in this regime we cannot
eliminate ``fast'' modes by Haldane's mapping. Further we will assume that
the ``renormalized classical'' regime (ii) takes place.

We use the same procedure as used by Haldane \cite{Haldane} (see full
discussion in Ref.\cite{ArovasBook}) to integrate out ``fast'' modes. Thus
the partition function has in terms of a path integral the form
\begin{eqnarray}
Z &=&\int D\mbox {\boldmath $\sigma $}_i(\tau )\exp \left\{ -\frac{\chi _0}2%
\int\limits_0^{1/T}d\tau \sum_i(\partial _\tau \mbox {\boldmath $\sigma $}%
_i)^2\right.  \nonumber \\
&&\left. -\frac 12S^2\int\limits_0^{1/T}d\tau \sum_{i\,j}J_{ij}(%
\mbox
{\boldmath $\sigma $}_i-\mbox {\boldmath $\sigma $}_j)^2\right\}
\prod\limits_i\delta (\mbox {\boldmath $\sigma $}_i^2-1)  \label{NLSM}
\end{eqnarray}
where $\mbox {\boldmath $\sigma $}$ is a three-component unit-length vector
field, $i$ is the index of a site, $\chi _0$ is the uniform magnetic
susceptibility. In the continual limit we reproduce the standard
three-dimensional quantum nonlinear-sigma model. However, in the quasi-2D
case the large value of $\xi _{J^{\prime }}$ gives a possibility to pass to
the continual limit only within the layers: $\mbox {\boldmath $\sigma $}%
_i(\tau )\rightarrow \mbox {\boldmath $\sigma $}_{i_z}({\bf r,\tau )}$ where
${\bf r}$ is a 2D vector, $i_z$ is the index of a layer. The partition
function takes the form
\begin{eqnarray}
Z &=&\int D\mbox {\boldmath $\sigma $}_{i_z}(r,\tau )\exp \left\{ -\frac{%
\rho _s^0}2\int\limits_0^{1/T}d\tau \int d^2{\bf r}\sum_{i_z}\left[ \frac 1{%
c_0^2}(\partial _\tau \mbox {\boldmath $\sigma $}_{i_z})^2\right. \right.
\nonumber \\
&&\left. \left. +({\bf \nabla }\mbox {\boldmath $\sigma $}_{i_z})^2+\frac
\alpha 2(\mbox {\boldmath $\sigma $}_{i_z+1}-\mbox {\boldmath $\sigma $}%
_{i_z})^2\right] \right\} \delta (\mbox {\boldmath $\sigma $}_{i_z}^2-1)
\end{eqnarray}
where $\rho _s^0=JS^2\ $is the bare spin stiffness, $c_0=(\rho _s^0/\chi
_0)^{1/2}$ is the bare value of the spin-wave velocity. Here and hereafter
we use the system of units where $a=1.$

To pass to the $O(N)$ model we replace the three-component field $%
\mbox
{\boldmath $\sigma $}_{i_z}({\bf r,\tau )}$ by the $N$-component one $\sigma
_{i_z}^m({\bf r},\tau )$, $m=1...N.$ The constraint condition $\sigma ^2=1$
may be taken into account by introducing the slave field $\lambda _{i_z}(%
{\bf r},\tau ).$ To calculate the dynamic susceptibility we also introduce
the external non-uniform time-dependent magnetic field $h_{i_z}^m({\bf r,}%
\tau ).$ Then we obtain the partion function of the $O(N)$ model in the
form:
\begin{eqnarray}
Z[h] &=&\int D\sigma D\lambda \exp \left\{ -\frac 1{2g}\int\limits_0^{1/T}d%
\tau \int d^2{\bf r}\sum_{i_z}\left[ \frac 1{c_0^2}(\partial _\tau \sigma
_{i_z})^2\right. \right.   \nonumber \\
&&\left. +({\bf \nabla }\sigma _{i_z})^2+\frac \alpha 2(\sigma
_{i_z+1}-\sigma _{i_z})^2+i\lambda (\sigma _{i_z}^2-1)\right.   \nonumber \\
&&\left. \left. -2gh_{i_z}(\sigma _{i_z}-\overline{\sigma })\right] \right\}
\label{zp}
\end{eqnarray}
where $g=N/\rho _s^0$ is the coupling constant, $\overline{\sigma }%
^m=\langle \sigma _{i_z}^m({\bf r},\tau )\rangle $ is the average part of
the field $\sigma $, which is supposed to be static and uniform. After
integrating over $\widetilde{\sigma }=\sigma -\overline{\sigma }$ the
partition function takes the form
\begin{eqnarray}
&&Z\,[h]
\begin{array}{c}
=
\end{array}
\int D\lambda \exp (NS_{eff}[\lambda ,h])  \label{Zef} \\
&&S_{eff}[\lambda ,h]
\begin{array}{c}
=
\end{array}
\frac 12\ln \det \widehat{G}_0+\frac 1{2g}(1-\overline{\sigma }^2)\text{Sp}%
(i\lambda )  \nonumber \\
&&\ +\frac 1{2g}\text{Sp}\left[ \left( i\lambda \overline{\sigma }-h/\rho
_s^0\right) \widehat{G}_0\left( i\lambda \overline{\sigma }-h/\rho
_s^0\right) \right]   \label{Sef}
\end{eqnarray}
where
\begin{eqnarray}
\widehat{G}_0 &=&[\partial _\tau ^2/c_0^2+{\bf \nabla }^2+\alpha \Delta
_z]^{-1} \\
\Delta _z\sigma _{i_z}({\bf r,\tau )} &=&\sigma _{i_z+1}({\bf r,\tau )}%
-\sigma _{i_z}({\bf r,\tau )}  \nonumber
\end{eqnarray}
Since $N$ enters (\ref{Zef}) only as a prefactor in the exponent, expanding
near the saddle point generates a series in $1/N$. At $T<T_{\text{Neel}}$ we
have the saddle point value $i\lambda =0$ and $\overline{\sigma }^2\neq 0$.
The Green's function of the field $\widetilde{\sigma }$ is defined by
\begin{eqnarray}
&&G^{mn}({\bf q},q_z,\omega _n)
\begin{array}{c}
=
\end{array}
\frac{\rho _s^0}{Z[0]}\int \frac{d^2{\bf p}}{(2\pi )^2}\int \frac{dp_z}{2\pi
}\sum_{\omega _l} \\
&&\left. \frac{\partial ^2Z[h]}{\partial h^m(p,p_z,\omega _l)\partial h^n(%
{\bf q-p},q_z-p_z,\omega _{l-n})}\right| _{h=0}  \nonumber
\end{eqnarray}
where $h({\bf p},p_z,\omega )$ is the Fourier transform of $h_{i_z}({\bf r}%
,\tau )$. Note that only diagonal elements $G^{mm}$ are nonzero, and they
are proportional to the non-uniform dynamic spin susceptibility:
\begin{equation}
G^{mn}({\bf q},q_z,\omega )=\frac{\rho _s^0}{S^2}\chi ^{mm}({\bf q+Q}%
,q_z+\pi ,\omega )\delta _{mn}  \label{ghi}
\end{equation}
where ${\bf Q=(}\pi ,\pi )$ is the wavevector of antiferromagnetic structure
in the plane; for $N=3$%
\begin{equation}
\chi ^{\alpha \beta }({\bf q},q_z,\omega )=\sum_ie^{i({\bf qR}%
_i+q_zR_i^z)}\langle \langle S_0^\alpha |S_i^\beta \rangle \rangle _\omega
\end{equation}
where $\,S_i^\alpha $ are spin operators, $\alpha ,\beta =x,y,z$. Since the
partition function $Z[0]$ is invariant under rotations in the spin space,
further we will assume $\overline{\sigma }^m=\overline{\sigma }\delta _{mN}$
where $\overline{\sigma }$ plays the role of the relative sublattice
magnetization $\overline{S}/S$. Then $G^{NN}$ corresponds to the
longitudinal Green's function, $G_l$, while other diagonal components (wich
are all equal) to the transverse Green's function, $G_t$. At $T<T_{\text{Neel%
}}$, the value of $\overline{\sigma }$ is determined by the constraint $%
\langle \sigma ^2\rangle =1$ which takes the form
\begin{equation}
1-\overline{\sigma }^2=\frac T{\rho _s^0}\sum_{\omega _n}\sum_m\int \frac{d^2%
{\bf k}}{(2\pi )^2}\int \frac{dk_z}{2\pi }G^{mm}(k,k_z,\omega _n)
\label{Constr0}
\end{equation}
We use the relativistic (hard) cutoff $\omega _n^2+k^2<\Lambda ^2$ of
frequency summations and momentum integrations; in this regularization
scheme the value of the bare spin wave velocity $c_0$ is replaced by the
fully renormalized one, $c$, which will be putted to be equal to unity
except for the final results.

In the limit $N\rightarrow \infty $ we may replace in (\ref{Sef}) $\lambda $
by its saddle-point value to obtain the ``free'' Green's function (which is
the same for transverse and longitudinal components)
\begin{equation}
G_0(k,k_z,\omega _n)=\left[ \omega _n^2+k^2+\alpha (1-\cos k_z)\right] ^{-1}
\end{equation}
After evaluation of the integrals and frequency summation in (\ref{Constr0})
we obtain the Neel temperature in the limit $N\rightarrow \infty $%
\begin{equation}
T_{\text{Neel}}^0=\frac{4\pi \rho _s^{N=\infty }}{N\ln (2T_{\text{Neel}%
}^2/\alpha c^2)}  \label{TNMF}
\end{equation}
where $\rho _s^{N=\infty }=N(1/g-1/g_c)$ is the renormalized spin stiffness
in zeroth order in $1/N,$ $g_c=2\pi ^2/\Lambda .$ To compare the result (\ref
{TNMF}) with the result of the SSWT we note that the value of spin stiffness
in SSWT is $\rho _s^{\text{SSWT}}=\gamma S\overline{S}_0$ (for $S=1/2$ this
equals to $0.176J$ which is somewhat lower than the result of two-loop RG
analysis \cite{Rs1} and numerical calculations \cite{Rs2}, $\rho _s=0.181J$)
and the value of the spin-wave velocity is $c^{\text{SSWT}}=\sqrt{8}\gamma S$%
. Thus we see that the value (\ref{TNMF}) is $N$ times smaller than the
corresponding SSWT value (\ref{TNSSWT}) (besides that, in SSWT $\alpha $ is
replaced by its renormalized value, $\alpha _c^{\text{SSWT}}=2\gamma
_c^{\prime }/\gamma _c<\alpha ).$ Further we will show that, as well as in
the calculation \cite{Chubukov} of the correlation length in the 2D-case in
the first order in $1/N,$

\noindent (i) the factor of $N$ in the denominator of (\ref{TNMF}) is to be
replaced by $N-2$

\noindent (ii) $\rho _s^{N=\infty }$ and $\alpha $ in (\ref{TNMF}) are to be
replaced by their renormalized values, $\rho _s$ and $\alpha _c$

\noindent (iii) terms of order of $\ln \ln (2T^2/\alpha )$ and unity, which
do not enter the SSWT result for $T_{\text{Neel}},$ occur in the denominator
of (\ref{TNMF}).

The exact Green's function may be expressed as
\begin{eqnarray}
G^{mm}(k,k_z,\omega _n) &=&\left[ \omega _n^2+k^2+\alpha (1-\cos k_z)\right.
\label{GrF} \\
&&\left. +\Sigma (k,k_z,\omega _n)\right] ^{-1}-C(k,k_z,\omega _n)\delta
_{mN}  \nonumber
\end{eqnarray}
To first order in $1/N$ the self-energy $\Sigma (k,k_z,\omega _n)$ and the
function $C(k,k_z,\omega _n),$ which describes renormalizations owing to the
long-range order, are given by \cite{Chubukov}
\begin{eqnarray}
\Sigma (k,k_z,\omega _n) &=&\frac{2T}N\sum_{\omega _m}\int \frac{d^2{\bf q}}{%
(2\pi )^2}\int\limits_{-\pi }^\pi \frac{dq_z}{2\pi }  \label{SE} \\
&&\ \times \ \frac{G_0({\bf k}+{\bf q},k_z+q_z,\omega _n+\omega
_m)-G_0(q,q_z,\omega _m)}{\widetilde{\Pi }(q,q_z,\omega _m)}  \nonumber \\
C(k,k_z,\omega _n) &=&\frac{2\overline{\sigma }^2}g\frac 1{\widetilde{\Pi }%
(k,k_z,\omega _n)}  \label{CE}
\end{eqnarray}
where
\begin{eqnarray}
\widetilde{\Pi }(q,q_z,\omega _n) &=&\Pi (q,q_z,\omega _n)+\frac 2g\overline{%
\sigma }^2G_0(q,q_z,\omega _n) \\
\Pi (q,q_z,\omega _n) &=&T\sum_{\omega _l}\int \frac{d^2{\bf p}}{(2\pi )^2}%
\int\limits_{-\pi }^\pi \frac{dp_z}{2\pi }G_0(p,p_z,\omega _l)  \nonumber \\
&&\ \ \times \ G_0({\bf p}+{\bf q},p_z+q_z,\omega _l+\omega _n).  \label{Pi}
\end{eqnarray}
Note that the quantity $C$ in (\ref{CE}) has in fact the zeroth order in $%
1/N,$ but the corresponding contribution to the constraint is of order of $%
1/N.$ The polarization operator $\Pi (q,q_z,\omega _n)$ determines the
longitudinal Green's function in the zeroth order in $1/N$%
\begin{equation}
G_l^{N=\infty }(q,q_z,\omega _n)=\frac{\Pi (q,q_z,\omega _n)}{q^2\Pi
(q,q_z,\omega _n)+2\overline{\sigma }^2/g}  \label{GL}
\end{equation}

To first order in $1/N$ the constraint (\ref{Constr0}) takes the form
\begin{eqnarray}
1-\overline{\sigma }^2 &=&gT\sum_{\omega _m}\int \frac{d^2{\bf k}}{(2\pi )^2}%
\int \frac{dk_z}{2\pi }G_0(k,k_z,\omega _m)  \label{Constr1/N} \\
&&-gT\sum_{\omega _m}\int \frac{d^2{\bf k}}{(2\pi )^2}\int \frac{dk_z}{2\pi }%
G_0^2(k,k_z,\omega _m)\Sigma (k,k_z,\omega _m)  \nonumber \\
&&-\frac{2\overline{\sigma }^2T}N\sum_{\omega _m}\int \frac{d^2{\bf k}}{%
(2\pi )^2}\int \frac{dk_z}{2\pi }\frac{G_0^2(k,k_z,\omega _m)}{\widetilde{%
\Pi }(k,k_z,\omega _m)}.  \nonumber
\end{eqnarray}
Following to \cite{Chubukov} we introduce the function
\begin{eqnarray}
I(k,k_z,\omega _m) &=&T\sum_{\omega _n}\int \frac{d^2{\bf q}}{(2\pi )^2}\int
\frac{dq_z}{2\pi }G_0^2(q,q_z,\omega _n)  \label{I} \\
&&\times \left[ G_0({\bf k}+{\bf q},k_z+q_z,\omega _m+\omega _n)\right.
\nonumber \\
&&-\left. G_0(k,k_z,\omega _m)\right]   \nonumber
\end{eqnarray}
and represent the equation (\ref{Constr1/N}) in the following convenient
form
\begin{eqnarray}
1 &=&gT\sum_{\omega _m}\int \frac{d^2{\bf k}}{(2\pi )^2}\int \frac{dk_z}{%
2\pi }G_0(k,k_z,\omega _m)  \nonumber \\
&&-gR(T,x_{\overline{\sigma }})+\overline{\sigma }^2\left[ 1-F(T,x_{%
\overline{\sigma }})\right]   \label{Constr1/Nm}
\end{eqnarray}
where
\begin{equation}
R(T,x_{\overline{\sigma }})=\frac{2T}N\sum_{\omega _m}\int \frac{d^2{\bf k}}{%
(2\pi )^2}\int \frac{dk_z}{2\pi }\frac{I(k,k_z,\omega _m)}{\widetilde{\Pi }%
(k,k_z,\omega _m)},  \label{R}
\end{equation}
\begin{equation}
F(T,x_{\overline{\sigma }})=\frac{2T}N\sum_{\omega _m}\int \frac{d^2{\bf k}}{%
(2\pi )^2}\int \frac{dk_z}{2\pi }\frac{G_0^2(k,k_z,\omega _m)}{\widetilde{%
\Pi }(k,k_z,\omega _m)},  \label{F}
\end{equation}
and
\begin{equation}
x_{\overline{\sigma }}=4\pi \overline{\sigma }^2/gT
\end{equation}
The calculation of functions $I,$ $\Pi $ for the quasi-2D case is presented
in Appendx A.

Thus the functions $R$ and $F$ determine the $1/N$-corrections to the
constraint. The expressions (\ref{Constr1/Nm})-(\ref{F}) enable one to
investigate the magnetization and to calculate the Neel temperature for a
quantum quasi-2D antiferromagnet.

\section{The sublattice magnetization, Neel temperature and correlation
functions}

As discussed in the beginning of previous section, we consider the quantum
case with $\alpha $ being small enough to satisfy the condition $\ln (2T_{%
\text{Neel}}^2/\alpha c^2)\gg 1.$ The calculation of the functions $R$ and $F
$ at $T\gg \alpha ^{1/2}$ (i.e. $T\gg (JJ^{\prime })^{1/2}$) is discussed in
Appendix B. Neglecting the terms of order of $1/\ln (2T_{\text{Neel}%
}^2/\alpha c^2)$ we have
\begin{eqnarray}
R(T,x_{\overline{\sigma }}) &=&\frac T{2\pi N}\ln \frac{2T^2}\alpha -\frac{%
(3+2x_{\overline{\sigma }})T}{4\pi N}\ln \frac{4\pi \rho _s}{NTx_{\overline{%
\sigma }}}  \nonumber \\
&&+\frac T{2\pi N}\frac{\ln (2T^2/\alpha )}{\ln (2T^2/\alpha )+x_{\overline{%
\sigma }}}  \nonumber \\
&&+\frac{8T}{3\pi ^3N}\ln \frac{2T^2}\alpha \ln \frac{N\Lambda }{16\rho _s}
\nonumber \\
&&-\frac{2T}{3\pi ^3N}\ln \frac{N\Lambda }{16\rho _s}+\frac T{4\pi }I_1(x_{%
\overline{\sigma }})  \label{R1}
\end{eqnarray}
and
\begin{equation}
F(T,x_{\overline{\sigma }})=\frac 1N\ln \frac{4\pi \rho _s}{NTx_{\overline{%
\sigma }}}+\frac 8{\pi ^2N}\ln \frac{N\Lambda }{16\rho _s}+I_2(x_{\overline{%
\sigma }})  \label{F1}
\end{equation}
where the functions $I_1(x_{\overline{\sigma }}),\,I_2(x_{\overline{\sigma }%
})$ are defined in Appendix B. After substituting (\ref{R1}), (\ref{F1})
into the constraint equation (\ref{Constr1/Nm}) and using the results of Ref.%
\cite{Chubukov} for the renormalized ground-state sublattice magnetization $%
\overline{\sigma }_0=\overline{\sigma }(T=0)=\overline{S}_0/S$ and the spin
stiffness $\rho _s$ of a quantum 2D antiferromagnet,
\begin{equation}
\frac{\overline{\sigma }_0^2}{\rho _s}=\frac gN\left( 1-\frac 8{3\pi ^2N}\ln
\frac{N\Lambda }{16\rho _s}\right) ,  \label{ms}
\end{equation}
\begin{equation}
\rho _s=\rho _s^{N=\infty }\left( 1+\frac{32}{3\pi ^2N}\ln \frac{N\Lambda }{%
16\rho _s}\right) ,  \label{rs}
\end{equation}
one can see that the sublattice magnetization, being expressed in terms of
quantum-renormalized $\rho _s$ and $\overline{\sigma }_0,$ still depends on $%
\Lambda ,$ i.e. is non-universal. To make the sublattice magnetization
completely universal we have to introduce the renormalized parameter of the
interlayer coupling
\begin{equation}
\alpha _r=\alpha \left[ 1-\frac 8{3\pi ^2N}\ln \frac{N\Lambda }{16\rho _s}%
\right] .  \label{AlfaR}
\end{equation}
We shall demonstrate below that at low enough temperatures any regular
(nondivergent) terms in the renormalized interlayer coupling parameter are
absent, so that this is renormalized only due to temperature fluctuations at
higher $T$. Being rewritten through the renormalized parameters, the
constraint equation (\ref{Constr1/Nm}) reads 
\begin{eqnarray}
&&1-\frac{NT}{4\pi \rho _s}\left[ (1-\frac 2N)\ln \frac{2T^2}{\alpha _r}+%
\frac 3N\ln \frac{4\pi \rho _s}{NTx_{\overline{\sigma }}}\right. 
\label{ConstrFF} \\
&&\left. -\frac 2N\frac{\ln (2T^2/\alpha _r)}{\ln (2T^2/\alpha _r)+x_{%
\overline{\sigma }}}-I_1(x_{\overline{\sigma }})\right]   \nonumber \\
\  &=&\frac{\overline{\sigma }^2}{\overline{\sigma }_0^2}\left[ 1+\frac 1N%
\ln \frac{4\pi \rho _s}{NTx_{\overline{\sigma }}}-I_2(x_{\overline{\sigma }%
})\right]   \nonumber
\end{eqnarray}
Note that we have simply replaced $\alpha $ by $\alpha _r$ in the terms of
order of $1/N$ in (\ref{ConstrFF}) since this yields an error of order of $%
1/N^2.$

First we consider the case $x_{\overline{\sigma }}\gg 1,$ or, equivalently, 
\begin{equation}
NT/4\pi \rho _s\ll \overline{\sigma }^2/\overline{\sigma }_0^2  \label{two-d}
\end{equation}
Since $x_{\overline{\sigma }}$ is the decreasing function of temperature,
this inequality is satified at low enough temperatures. In this case the
integrals $I_1(x_{\overline{\sigma }})$ and $I_2(x_{\overline{\sigma }})$
are of order of $1/x_{\overline{\sigma }},$ i.e. are small. Thus to leading
(zeroth) order in $1/x_\sigma $ the constraint equation (\ref{ConstrFF})
coincides with that in the case of space dimensionality $d=2+\varepsilon $
(Appendix C) with the replacement $1/\varepsilon \rightarrow \ln (2/\alpha
), $ which corresponds to the limit $\varepsilon \rightarrow 0$ with
simultaneous cutting the integrals over quasimomentum on the scale $1/\xi
_{J^{\prime }}.$ Similar to the $d=2+\varepsilon $ case (Appendix C) we
transform the logarithmic term in the right-hand side of (\ref{ConstrFF})
into power and replace $N\rightarrow N-2.$ Then we have 
\begin{eqnarray}
&&(\overline{\sigma }/\overline{\sigma }_0)^{1/\beta _2}\left[ 1-I_2(x_{%
\overline{\sigma }})\right]  \label{Constr2D} \\
&=&1-\frac{NT}{4\pi \rho _s}\left[ (1-\frac 2N)\ln \frac{2T^2}{\alpha _r}+%
\frac 3N\ln \frac{\overline{\sigma }_0^2}{\overline{\sigma }^2}\right. 
\nonumber \\
&&\left. -\frac 2N\frac{\ln (2T^2/\alpha _r)}{\ln (2T^2/\alpha _r)+x_{%
\overline{\sigma }}}-I_1(x_{\overline{\sigma }})\right]  \nonumber
\end{eqnarray}
where, being expressed through the renormalized parameters, 
\begin{equation}
x_{\overline{\sigma }}=\frac{4\pi \rho _s}{(N-2)T}\frac{\overline{\sigma }^2%
}{\overline{\sigma }_0^2}
\end{equation}
The ``critical exponent'' $\beta _2,$ which is the limit of $\beta
_{2+\varepsilon }$ at $\varepsilon \rightarrow 0,$ is given by 
\begin{equation}
\beta _2=\frac 12\frac{N-1}{N-2}  \label{b2}
\end{equation}

As well as in the $d=2+\varepsilon $ case, \ two-regimes are possible under
the condition (\ref{two-d}). Consider first the low-temperature (spin-wave)
region where 
\begin{equation}
T(N-2)\ln (2T^2/\alpha )/4\pi \rho _s\ll \overline{\sigma }^2/\overline{%
\sigma }_0^2  \label{isw}
\end{equation}
In this region 
\begin{equation}
\overline{\sigma }=\overline{\sigma }_0\left[ 1-\frac{T(N-1)}{8\pi \rho _s}%
\ln \frac{2T^2}{\alpha _rc^2}\right] .  \label{MagnLowT}
\end{equation}
At $N=3$ we reproduce the result of SSWT (\ref{MagnSSWT}) with $2\gamma
^{\prime }/\gamma $ being replaced by $\alpha _r$. The factor $N-1$ in (\ref
{MagnLowT}) has a simple physical meaning: this is the number of gapless
(Goldstone) modes. We can conclude that in the temperature interval (\ref
{isw}) spin-wave excitations give the main contribution to the dependence $%
\overline{\sigma }(T)$. Note that the spin-wave result (\ref{MagnLowT}) can
be obtained also from the untransformed constraint equation (\ref{ConstrFF}).

To demonstrate that in the interval (\ref{isw}) the experimentally observed
interlayer exchange parameter coincides with $\alpha _r$ we calculate the
self-energy $\Sigma (k,k_z,0).$ By using (\ref{SE}) we get 
\begin{equation}
\Sigma (k,k_z,0)=\frac{8k^2}{3\pi ^2N}\ln \frac{N\Lambda }{16\rho _s}
\end{equation}
irrespectively of $k_z.$ Thus we have 
\begin{eqnarray}
G_t^{-1}(k,k_z,0) &=&k^2\left[ 1+\frac 8{3\pi ^2N}\ln \frac{N\Lambda }{%
16\rho _s}\right] +\alpha (1-\cos k_z) \\
&=&Z^{-1}\left[ k^2+\alpha _r(1-\cos q_z)\right] .  \nonumber
\end{eqnarray}
We see that the renormalized Green's function differs from the bare one by
the renormalization factor $Z$ and by replacement $\alpha \rightarrow \alpha
_r$ only. Thus the experimentally observed (fully renormalized) interlayer
coupling is just $\alpha _r.$ At higher temperatures the temperature
renormalization of interlayer coupling, which will be calculated below,
becomes important.

At intermediate temperatures where 
\begin{equation}
(N-2)T/4\pi \rho _s\ll \overline{\sigma }^2/\overline{\sigma }_0^2\ll
(N-2)T\ln (2T^2/\alpha )/4\pi \rho _s  \label{isp}
\end{equation}
we have a 2D-like critical behavior of the sublattice magnetization, 
\begin{equation}
(\overline{\sigma }/\overline{\sigma }_0)^{1/\beta _2}=1-\frac T{4\pi \rho _s%
}\left[ (N-2)\ln \frac{2T^2}{\alpha _r}+3\ln \frac{\overline{\sigma }_0^2}{%
\overline{\sigma }^2}-2\right]  \label{Magn2D}
\end{equation}
For $N=3$ we have $\beta _2=1,$ which coincides with the critical exponent
of SWT and SSWT. However, the term with $\ln (\overline{\sigma }^2/\overline{%
\sigma }_0^2),$ which is present in (\ref{Magn2D}), leads to a significant
modification of the dependence $\overline{\sigma }(T)$ in the temperature
region under consideration in comparison with SSWT and leads to a
considerable lowering the Neel temperature. With further approaching the
transition point the behavior of the order parameter changes to the 3D one.

Consider the temperatures which are very close to $T_{\text{Neel}}$, so that 
$\overline{\sigma }$ is small enough to satisfy the inequality $x_{\overline{%
\sigma }}\ll 1,$ i.e. 
\begin{equation}
\overline{\sigma }^2/\overline{\sigma }_0^2\ll (N-2)T/4\pi \rho _s
\label{isc}
\end{equation}
After expanding (\ref{ConstrFF}) near $T=T_{\text{Neel}},$ $x_{\overline{%
\sigma }}=0,$ picking out the logarithmically divergent parts of $I_1(x_{%
\overline{\sigma }})$ and $I_2(x_{\overline{\sigma }})$ at small $x_{%
\overline{\sigma }}$ analytically, and evaluating numerically the integrals
we have 
\begin{eqnarray}
1-\frac T{T_{\text{Neel}}} &=&\frac{\overline{\sigma }^2}{\overline{\sigma }%
_0^2}\left[ 1+\frac 1N\ln \frac{4\pi \rho _s}{(N-2)T_{\text{Neel}}}\right. 
\label{MagnCr} \\
&&\left. +\frac 8{\pi ^2N}\ln x_{\overline{\sigma }}-A_0\right]   \nonumber
\end{eqnarray}
where $A_0=2.8906/N$. The equation for $T_{\text{Neel}}$ reads 
\begin{eqnarray}
T_{\text{Neel}} &=&4\pi \rho _s\left[ (N-2)\ln \frac{2T_{\text{Neel}}^2}{%
\alpha _r}\right.   \label{TN1} \\
&&\left. +3\ln \frac{4\pi \rho _s}{(N-2)T_{\text{Neel}}}-0.0660\right] ^{-1}.
\nonumber
\end{eqnarray}
As will be clear below the second term in the denominator, which is of order
of $\ln \ln (2T_{\text{Neel}}^2/\alpha )$ leads to a significant lowering of
Neel temperature in comparison with SSWT (where only the first term is taken
into account). We collect separately the logarithmic terms in (\ref{MagnCr})
which comes from the quasimomenta $q\gg \alpha ^{1/2}$ (2D regime) and $q\ll
\alpha ^{1/2}$ (3D regime): 
\begin{eqnarray}
1-\frac T{T_{\text{Neel}}} &=&\frac{\overline{\sigma }^2}{\overline{\sigma }%
_0^2}(1-A_0)\left[ 1+\frac 1N\ln \frac{4\pi \rho _s}{(N-2)T_{\text{Neel}}}%
\right]  \\
&&\times \left[ 1+\frac 8{\pi ^2N}\ln x_{\overline{\sigma }}\right]  
\nonumber
\end{eqnarray}
Unlike the ``2D-like'' regime, the coefficients at the logarithms are
different. Transforming the logarithmic terms into powers we obtain 
\begin{equation}
\frac{\overline{\sigma }^2}{\overline{\sigma }_0^2}=\left[ \frac{4\pi \rho _s%
}{(N-2)T_{\text{Neel}}}\right] ^{\beta _3/\beta _2-1}\left[ \frac 1{1-A_0}%
\left( 1-\frac T{T_{\text{Neel}}}\right) \right] ^{2\beta _3}
\label{MagnCr1}
\end{equation}
where 
\begin{equation}
\beta _3=\frac 12\left( 1-\frac 8{\pi ^2N}\right) .  \label{b3}
\end{equation}
is the true 3D critical exponent for the magnetization. It should be noted
that we have not to perform the replacement $N\rightarrow N-2$ in (\ref{b3})
and other contributions which come from essentially three-dimensional
integrals. We get for $N=3$ the value $\beta _3\simeq 0.36.$ The result (\ref
{b3}) coincides with that of the $1/N$ expansion in the $\phi ^4$ model \cite
{Ma} at $d=3$, in agreement with the universality hypothesis. The dependence
(\ref{MagnCr1}) is to be compared with that in the Tyablikov approximation (%
\ref{MagnTbT}) where $\beta =1/2$ and the dimensional crossover is absent.

Consider now the self-energy $\Sigma (k,k_z,0)$ at $T=T_{\text{Neel}}.$ At $%
\alpha ^{1/2}\ll k\ll T_{\text{Neel}}$ the self-energy has the same form as
in the 2D case \cite{Chubukov} with $\xi $ being replacing by $\xi
_{J^{\prime }}$: 
\begin{equation}
\Sigma (k,k_z,0)=k^2\left[ \eta \ln \frac{N\Lambda }{16\rho _s}+\frac 1N\ln 
\frac{\ln (2T_{\text{Neel}}^2/\alpha )}{\ln (2k^2/\alpha )}+\frac 1N\right] .
\end{equation}
Thus the expression for Green's function reads ($G=G_t=G_l$) 
\begin{eqnarray}
G(k,k_z,0) &=&\frac 1{k^2}\left[ \frac{(N-2)T_{\text{Neel}}}{4\pi \rho _s}%
\ln \frac{2k^2}\alpha \right] ^{1/(N-2)}  \label{GreenC2} \\
&&\ \times \frac{N-1}N\left[ 1-\eta \ln \frac{N\Lambda }{16\rho _s}\right] 
\nonumber \\
&&\ \;\;\;\;\;\;\;\;\;\;\;\;\;\; 
\begin{array}{c}
\alpha ^{1/2}\ll k\ll T_{\text{Neel}}
\end{array}
\nonumber
\end{eqnarray}
At $k\ll \alpha ^{1/2}$ , $k_z\ll 1$ the $k$-dependence of the Green's
function changes. After integration and frequency summation in (\ref{SE})\
(which are analogous to the calculation of the functions $R$ and $F$ in
Appendix B) we have 
\begin{eqnarray}
\Sigma (k,k_z,0) &=&A_1k^2+\frac \alpha 2A_2k_z^2  \label{Sigma3D} \\
&&\ +\frac \eta 2(k^2+\frac \alpha 2k_z^2)\ln \frac \alpha {k^2+\alpha
k_z^2/2}.  \nonumber
\end{eqnarray}
Here 
\begin{eqnarray}
A_1 &=&\eta \ln \frac{N\Lambda }{16\rho _s}+\frac 1N\ln \ln \frac{2T^2}\alpha
+\frac{0.4564}N,  \label{A1A2} \\
\,\,A_2 &=&-0.6122/N.  \nonumber
\end{eqnarray}
and 
\begin{equation}
\eta =8/(3\pi ^2N)
\end{equation}
is the 3D critical exponent for the asymptotics of the correlation function
at the phase transition point in the first order in $1/N.$ For $N=3\,$we
have $\eta \simeq 0.09.$ Using (\ref{Sigma3D}) we find

\begin{eqnarray}
G^{-1}(k,k_z,0) &=&(1+A_1)\alpha _c^{\eta /2}\left( k^2+\frac{\alpha _c}2%
k_z^2\right) ^{1-\eta /2}  \label{GreenC} \\
&&\ \;\;\;\;\;\;\;\;\;\;\;\;\;\;\;\; 
\begin{array}{c}
k\ll \alpha ^{1/2},k_z\ll 1
\end{array}
.  \nonumber
\end{eqnarray}
The quantity 
\begin{equation}
\alpha _c=\alpha (1+A_2)/(1+A_1)  \label{AlfaC}
\end{equation}
can be interpreted as the renormalized interlayer coupling at $T=T_{\text{%
Neel}}.$

Using (\ref{AlfaR}) we find the following relation between the renormalized
coupling parameters at low $T$ and at $T=T_{\text{Neel}}$: 
\begin{equation}
\alpha _c=\alpha _r\left( 1+\frac{1.0686}N\right) \left[ \frac{(N-2)T_{\text{%
Neel}}}{4\pi \rho _s}\right] ^{1/(N-2)}.  \label{ac}
\end{equation}
When deriving (\ref{ac}) we have transformed the term with $\ln \ln
(2T^2/\alpha )$ into a power and then replaced $N\rightarrow N-2$ in the
exponent. As well as in SSWT (see Sect.2), the renormalized interlayer
coupling at $T_{\text{Neel}}$ is lower than the low-temperature one, but the
concrete expression at $N=3$ is slightly different from these in SSWT.

Using (\ref{ac}) we get the following equation for $T_{\text{Neel}}$ in
terms of $\alpha _c$ 
\begin{eqnarray}
T_{\text{Neel}} &=&4\pi \rho _s\left[ (N-2)\ln \frac{2T_{\text{Neel}}^2}{%
c^2\alpha _c}\right.   \label{TN2} \\
&&\ \ \left. +2\ln \frac{4\pi \rho _s}{(N-2)T_{\text{Neel}}}+1.0117\right]
^{-1}.  \nonumber
\end{eqnarray}
where $c$ is the fully-renormalized spin-wave velocity; in SSWT we have $c=%
\sqrt{8}\gamma (0)S$ (see Sect.2). For $N=3$ we have 
\begin{equation}
T_{\text{Neel}}=4\pi \rho _s/\ln \left[ 5.5005\frac{(4\pi \rho _s)^2}{%
c^2\alpha _c}\right]   \label{TNT}
\end{equation}
which is similar to the result of the Tyablikov approximation (\ref{TNTYAB}%
), but the bare value of $\alpha $ is replaced by its renormalized value at
the critical temperature (\ref{ac}) and $\rho _s$ is also replaced by its
renormalized value. Besides that, the result (\ref{TNT}) does not violate
the scaling form (\ref{TNSCAL}).

Finally, we consider the spin correlation function 
\begin{eqnarray}
S(R,R_z) &=&-\frac 1\pi \int \frac{d^2{\bf k}}{(2\pi )^2}\int\limits_{-\pi
}^\pi \frac{dk_z}{2\pi }e^{i({\bf kR+}k_zR_z)} \\
&&\ \times \sum_m\int d\omega \,\text{Im\thinspace }\chi ^{mm}(k,k_z,\omega )%
\frac 1{e^{\omega /T}-1}  \nonumber
\end{eqnarray}
at $T=T_{\text{Neel}}$. For $N=3$ we have 
\[
S(R,R_z)=|<{\bf S}_i({\bf r}){\bf S}_{i+R_z}({\bf r}+{\bf R})>|
\]
The static approximation is sufficient to determine the asymptotics of the
correlation function. Using (\ref{ghi}), (\ref{ms}) we derive (cf. Ref.\cite
{Chubukov}) 
\begin{eqnarray}
S(R,R_z) &=&\frac{T\overline{S}_0^2}{\rho _s}\left[ 1+\eta \ln \frac{%
N\Lambda }{16\rho _s}\right]   \label{Corr} \\
&&\times \int \frac{d^2{\bf k}}{(2\pi )^2}\int\limits_{-\pi }^\pi \frac{dk_z%
}{2\pi }G(k,k_z,0)e^{i({\bf kR+}k_zR_z)}.  \nonumber
\end{eqnarray}
One can see that at $R^2\alpha +R_z^2\gg 1$ the asymptotics of the integral
in (\ref{Corr}) is determined by the region $k\ll \alpha ^{1/2}$ and $k_z\ll
1$ where $G(k,k_z,0)$ is calculated above (see (\ref{GreenC})). Substituting
(\ref{GreenC}) into (\ref{Corr}) we have 
\begin{eqnarray}
S({\cal R}) &=&\frac 1{4\pi }\frac{T_{\text{Neel}}\overline{S}_0^2}{\rho _s}%
(1-A_1-\eta C)\left[ 1+\eta \ln \frac{N\Lambda }{16\rho _s}\right]  
\nonumber \\
&&\times \left( \frac 2{\alpha _c^{1+\eta }{\cal R}^{2+2\eta }}\right) ^{1/2}
\end{eqnarray}
where ${\cal R}=(R^2+2R_z^2/\alpha _c)^{1/2}$ and $C\simeq 0.5772$ is the
Euler constant. Using the value of $A_1$ (\ref{A1A2}) and transforming the
term with $1/N\ln \ln (2T^2/\alpha )$ into power we obtain the final result
for the spin correlation function at $T=T_{\text{Neel}}$ and ${\cal R}\alpha
_c^{1/2}\gg 1:$%
\begin{eqnarray}
S({\cal R}) &=&\frac{T_{\text{Neel}}\overline{S}_0^2}{4\pi \rho _s}\left[ 
\frac{(N-2)T_{\text{Neel}}}{4\pi \rho _s}\right] ^{1/(N-2)}  \nonumber \\
&&\times (1-\overline{A}_1-\eta C)\left( \frac 2{\alpha _c^{1+\eta }{\cal R}%
^{2+2\eta }}\right) ^{1/2}
\end{eqnarray}
where $\overline{A}_1=0.4564/N.$ Thus $S({\cal R})$ enables one to determine
the value of $\alpha _c.$ As one should expect, being rewritten through
renormalized parameters $\rho _s$ and $\alpha _c,$ $S({\cal R})$ does not
contain the cutoff parameter $\Lambda $ and is thereby completely universal.

At $1\ll R\ll \alpha _c^{1/2}$ we derive from (\ref{GreenC2}), (\ref{Corr})
the leading term of asymptotics of the correlation function within a plane ($%
N=3$) 
\begin{equation}
S(R,0)=\frac{\overline{S}_0^2}{3\rho _s}\left( \frac T{4\pi \rho _s}\right)
^2\ln ^2\frac 8{\alpha _cR^2}
\end{equation}
Thus we have in this case a logarithmic decrease of the correlation
function, as well as in the 2D case at $1\ll R\ll \xi $ \cite{Chakraverty}.

\section{Discussion and conclusions}

In the above treatment we analyzed the sublattice magnetization $\overline{S}
$ of a quasi-2D quantum antiferromagnet ($T_{\text{Neel}}\ll JS$). At
temperatures $T\leq $ $(JJ^{\prime })^{1/2}$ the behavior $\overline{S}(T)$
is satisfactory decribed by the standard spin-wave theory. For $T\gg
(JJ^{\prime })^{1/2}$ we have obtained the equation (\ref{ConstrFF}) which
determines $\overline{S}$ to first order in the formal small parameter $1/N$%
. We have three temperature intervals (the boundaries of the intervals are
presented for $N=3$)

\noindent (i) the case of low temperatures 
\begin{equation}
T\ln (2T^2/JJ^{\prime })/(4\pi JS^2)\ll \overline{S}^2/\overline{S}_0^2
\end{equation}
[$\overline{S}_0=\overline{S}(T=0);$ $\overline{S}_0=S-0.196$ for the square
lattice] where the results of SSWT are reproduced.

\noindent (ii) the case of the intermediate temperatures (\ref{isp}), or
equivalently 
\begin{eqnarray}
\overline{S}^2/\overline{S}_0^2 &\ll &T\ln (2T^2/JJ^{\prime })/(4\pi JS^2) 
\nonumber \\
1-T/T_{\text{Neel}} &\gg &(1-A_0)(T/4\pi JS^2)^{1/2}
\end{eqnarray}
($1-A_0\simeq 0.0365$), where a 2D-like critical behavior, which is similar
to that in SSWT, takes place. However, the corrections to SSWT modify
considerably the numerical factors, so that the Neel temperature is
considerably lowered

\noindent (iii) the vicinity of the Neel temperature (\ref{isc}), or 
\begin{equation}
1-T/T_{\text{Neel}}\ll (1-A_0)(T/4\pi JS^2)^{1/2}
\end{equation}
where we obtain the critical behaviour $\overline{S}\sim (T_{\text{Neel}%
}-T)^{\beta _3}\,,$ $\beta _3\simeq 0.36$.

\noindent The detailed description of the temperature region between (ii)
and (iii), where $\overline{S}^2/\overline{S}_0^2\sim T/4\pi \rho _s,$
cannot be obtained within the first order in $1/N,$ since the equation (\ref
{ConstrFF}) is transformed in different ways in these regions to derive the
results (\ref{Constr2D}) and (\ref{MagnCr}) respectively. Note, that in the
region (ii) the ``2D-like'' behavior of the system enables one to calculate
corrections to SSWT in a regular way, e.g., by using the $1/N$-expansion in
the $CP^{N-1}$ model.

We have also derived the expressions for the magnetic transition temperature
(\ref{TN1}), (\ref{TN2}) which contain the renormalized quantities $\alpha
_{r,c}=2\gamma _{r,c}^{\prime }/\gamma $ where $\gamma _{r,c}^{\prime }$ are
the experimentally observable (renormalized) interlayer exchange parameters
at low temperatures and $T=T_{\text{Neel}}$ respectively, and $\gamma \simeq
1.1571J$ is the value of renormalized intralayer coupling parameter which is
weakly temperature dependent. Therefore these expressions have an universal
form, in agreement with the scaling analysis. Unlike the corresponding
results of the spin-wave approaches (see Sect.2), they contain the terms of
order of $\ln \ln (T_{\text{Neel}}^2/JJ^{\prime })$ and unity, which are
formally small as compared to the leading term of order of $\ln (T_{\text{%
Neel}}^2/JJ^{\prime })$. However, the lnln-terms result in a significant
lowering of $T_{\text{Neel}}$ in comparison with the SSWT value (\ref
{MagnSSWT}) at not too large $\ln (J/J^{\prime }).$ The regular terms yield
small corrections only, so that one may expect that the higher-order terms
in $1/\ln (T_{\text{Neel}}^2/JJ^{\prime })$ may be neglected.

The experimental temperature dependence \cite{Keimer} of the sublattice
magnetization in La$_2$CuO$_4$ is shown in Fig.1. For comparison, the
results of spin-wave approximations (SWT, SSWT and the Tyablikov theory, see
Sect.2) and the result of $1/N$ expansion are also presented. The
renormalized value of the in-plane exchange parameter $\gamma \simeq 1850K$
can be found from the experimental data \cite{Veloc} and the value $\gamma
_r^{\prime }/\gamma =5\cdot 10^{-4}$ was chosen from the best fit to SWT at
low temperatures $T<100$ K). The experimental results for $\gamma ^{\prime
}/\gamma $ are not reliable, and it is difficult to compare our value of $%
\gamma ^{\prime }/\gamma $ with experiment. For example, the result of Ref.%
\cite{Anis}, $\gamma ^{\prime }/\gamma =5\cdot 10^{-5},$ is by an order
lower than that found from the fit in the spin-wave region. It is also
important that the value of $\gamma ^{\prime }/\gamma $ has an appreciable
temperature dependence because of renormalizations. In particular, we have
from (\ref{ac}) for above parameters $\alpha _c/\alpha _r=\gamma _c^{\prime
}/\gamma _r^{\prime }\simeq 0.13.$ Thus experiments at different
temperatures may give different results.

One can see that SWT and SSWT yield satisfactory results for $T<0.6T_{\text{%
Neel}}$ and $T<0.8T_{\text{Neel}}$ respectively. At higher temperatures the
sublattice magnetization in SWT and SSWT is still linear in temperature, so
that the critical exponent is $\beta _{\text{SW}}=1,$ instead of the
experimental one, $\beta _{\text{exp}}\simeq 0.33.$ Besides that, both
theories give large values of the Neel temperature $T_{\text{Neel}}^{\text{%
SWT}}=672$K$,T_{\text{Neel}}^{\text{SSWT}}=537$K. This fact is often not
taken into account when treating experimental data. At the same time, TT
gives the value $T_{\text{Neel}}^{\text{TT}}=454$ K which is much lower than
those in SWT and SSWT and the magnetization critical exponent $\beta _{\text{%
TT}}=1/2.$ Thus the Tyablikov approximation seems to describe the
experimental data more satisfactorily. However, this approximation may be
justified in fact only in the case of ``classical'' magnets with $T_{\text{%
Neel}}\gg JS.$ Besides that, TT has a number of drawbacks: mean-field values
of critical exponents, absence of crossover from 2D-like to 3D behavior of
magnetization, neglect of quantum effects at high temperatures (in
particular, $T_{\text{Neel}}=T_{\text{Curie}}$ for the same $|\,J\,|$).

The result of the $1/N$-expansion to first order in $1/N$ (\ref{TN1}) is $T_{%
\text{Neel}}=345$ K which is considerably lower than in the Tyablikov
approximation and is in a good agreement with the experimental value, $T_{%
\text{Neel}}^{\exp }=325$K. The spin-wave region extends up to $300K,$ while
the 2D-like region from $320K$ to about $340K;$ the critical 3D region is
narrow (about $1K$)$.$ The results of the numerical solution of equation (%
\ref{Constr2D}) in the temperature regions (i) and (ii) and the dependence (%
\ref{MagnCr}) in the region (iii) turn out to be smoothly joined. One can
see also that the result of the $1/N$-expansion is most close to the
experimental data and demonstrates a correct critical behavior. One may
assume that higher-order $1/N$-corrections will give a precise description
of the experimental situation. Thus we may conclude that using the $1/N$%
-expansion in the $O(N)$ model improves considerably the results of standard
spin-wave approximations in the Heisenberg model.

Recent experiments demonstrate existence of a gap for the out-of-plane
spin-wave excitations\cite{Keimer} in La$_2$CuO$_4,$ which is assumed to be
determined by the easy-plane anisotropy. The possible role of easy-axis
anisotropy was also discussed, see, e.g., Ref.\cite{Easy-ax}. Therefore an
extension of the present approach to 2D systems with a weak anisotropy is of
interest. The results may be expected to be similar to those in the quasi-2D
case, since SSWT gives similar descriptions of both the types of magnets
with small ordering temperature \cite{OurFMM}.

The case of ``classical'' spins cannot be treated consistently in the
continual limit since in this case the natural upper limit cutoff parameter
(which is the temperature in the quantum case) is absent, and the integrals
are determined by the whole Brillouin zone. Therefore the continual models
may be used to calculate the critical exponents, but not the temperature
dependence of magnetization in a broad interval and the Neel temperature.

It would be also interesting to perform similar calculations of
thermodynamic properties for a ferromagnet. The results should coincide with
those for a antiferromagnet only in the classical case. Unfortunately, the
nonlinear-sigma model for ferromagnet has the Berry phase term ${\bf A(}%
\mbox {\boldmath $\sigma $}{\bf )}\partial \mbox {\boldmath $\sigma $}%
/\partial \tau $ in the action (${\bf A}$ is the vector potential of unit
magnetic monopole), see, e.g., Ref.\cite{ArovasBook}. This term cannot be
eliminated in the quantum case and prevents constructing the $1/N$%
-expansion. For singular contributions, the results for a quantum
ferromagnet ($T_{\text{Curie}}\ll JS)$ differ from those for an
antiferromagnet by the replacement 
\begin{equation}
\ln (T^2/8S^2\gamma \gamma _{r,c}^{\prime })\rightarrow \ln (T/\gamma
_{r,c}^{\prime }S)
\end{equation}
(as well as in SSWT, see Ref.\cite{OurFMM}). Taking into account only such
terms, the expression for the Curie temperature has the form 
\begin{equation}
T_{\text{Curie}}=4\pi \rho _s\left[ (N-2)\ln \frac{T_{\text{Curie}}}{\gamma
_r^{\prime }S}+3\ln \frac{2\pi \rho _s}{(N-2)T_{\text{Curie}}}+{\cal O}%
(1)\right] ^{-1}.
\end{equation}
or, in terms of the renormalized exchange parameter at the Curie
temperature, 
\begin{equation}
T_{\text{Curie}}=4\pi \rho _s\left[ (N-2)\ln \frac{T_{\text{Curie}}}{\gamma
_c^{\prime }S}+2\ln \frac{2\pi \rho _s}{(N-2)T_{\text{Curie}}}+{\cal O}%
(1)\right] ^{-1}
\end{equation}
where 
\begin{equation}
\gamma _c^{\prime }=A_\gamma \gamma _r^{\prime }\left[ \frac{(N-2)T_{\text{%
Curie}}}{4\pi \rho _s}\right] ^{1/(N-2)}.
\end{equation}
and $A_\gamma \sim 1.$ One may expect that, as well as in (\ref{TN1}), (\ref
{TN2}) the non-singular terms will influence weakly the value of the
ordering point. These regular contributions may be calculated for a
ferromagnet within the $1/N$-expansion in the $SU(N)$ model (cf.\cite
{ArovasBook}). However, as discussed in the Introduction, this expansion
gives poor results at not too large $N$ for $d$ not too close to 2$,$ so
that only the ``2D-like'' region can be described satisfactory. The
description of the 3D critical behavior of a quasi-2D ferromagnet requires
other methods.

We are grateful to M.I.Katsnelson for stimulating discussions.

\section*{Appendix A. Analytical results for the functions $\Pi (\lowercase {%
q,q_z,\omega _n})$ and $I(\lowercase {q,q_z,\omega _n})$}

Here we present a list of results for the polarization operator $\Pi
(q,q_z,\omega _n)$ (\ref{Pi}) and the function $I(q,q_z,\omega _n)$
determined by (\ref{I}) at $\alpha ^{1/2}\ll T$ , and the asymptotic forms
of these functions.

Due to inequality $\alpha ^{1/2}\ll T,$ the possible values of $q,\omega _n$
may be divided into two regions. The first region is $\omega _n=0,$ $q\ll T,$
while in the second region $q^2+\omega _n^2\gg \alpha ,$ i.e. either $\omega
_n=0$ and $q\gg \alpha ^{1/2}$ or $\omega _n\neq 0$ at arbitrary $q.$

It may be checked that at $q\ll T$ and $\omega _n=0$ the main contribution
to $\Pi $ comes from the term with $\omega _m=0.$ After integrating over $%
{\bf k,}k_z$ with the use of the Feynman identity (see, e.g., Ref.\cite
{Ryder}) we get 
\begin{eqnarray}
\Pi (q,q_z,0) &=&\frac{TK_2}2\int\limits_0^1\frac{dz}{\sqrt{z(1-z)}}\frac 1{%
\sqrt{z(1-z)q^4+2\widetilde{\alpha }(q,q_z)}},  \label{Pi1} \\
&&\
\,\,\,\,\,\,\,\,\,\,\,\,\,\,\,\,\,\,\,\,\,\,\,\,\,\,\,\,\,\,\,\,\,\,\,\,\,\,%
\,\,\,\,\,\,\,\,\,\,\,\,\,\,\,\,\,\,\,\,\,\,\,\,\,\,\,\,\,\,\,\,\,\,\,\,\,\,%
\,\,\,\,\,\,\,\,\,\,\,\,\,\,\,\,\,\,\,\,\,\,\,\,\,\,\,\, 
\begin{array}{c}
q\ll T
\end{array}
\nonumber
\end{eqnarray}
where $\widetilde{\alpha }(q,q_z)=\alpha [q^2+\alpha (1-\cos q_z)].$ At
large $q\gg \alpha ^{1/2}$ the function $\Pi (q,q_z,0)$ has a $2D$ form (cf.%
\cite{Chubukov}) 
\begin{equation}
\Pi (q,q_z,0)\simeq \frac T{2\pi q^2}\ln \frac{2q^2}\alpha
,\,\,\,\,\,\,\alpha \ll \,q^2\ll T.  \label{Pi1a1}
\end{equation}
In the opposite limit the form of function $\Pi (q,q_z,0)$ changes from $2D$
to $3D$ one:

\begin{equation}
\Pi (q,q_z,0)\simeq \frac T{4\sqrt{2\widetilde{\alpha }(q,q_z)}}%
,\,\,\,\,\,\,\,\,\,q^2\ll \alpha .  \label{Pi1a2}
\end{equation}

Consider now the case $q^2+\omega _n^2\gg \alpha .$ Picking out the terms
with $m=0$ and $m=-n$ (if $n\neq 0$) we have 
\begin{eqnarray}
\Pi (q,q_z,\omega _n) &=&\frac T{2\pi (q^2+\omega _n^2)}\ln \frac{%
2(q^2+\omega _n^2)^2}{q^2\alpha }+\Pi _{\text{qu}}(q,\omega _n),\,
\label{Pi2} \\
&&\ \ \ \,\,\,\,\ \qquad \qquad \qquad \qquad \qquad \qquad 
\begin{array}{c}
q^2+\omega _n^2\gg \alpha
\end{array}
\nonumber
\end{eqnarray}
where the quantum contribution $\Pi _{\text{qu}}$ is given by 
\begin{eqnarray}
\Pi _{\text{qu}}(q,\omega _n) &=&\frac T\pi \sum_{m\neq 0}\frac 1{\sqrt{%
(\omega _n^2+q^2+2\omega _m\omega _n)^2+4q^2\omega _m^2}} \\
&&\ \times \text{Arc}\tanh \frac{\omega _n^2+q^2+2\omega _m\omega _n}{\sqrt{%
(\omega _n^2+q^2+2\omega _m\omega _n)^2+4q^2\omega _m^2}}  \nonumber
\end{eqnarray}
In all the further calculations we will need only the asymptotic form of $%
\Pi _{\text{qu}}(q,\omega _n)$ for $q^2+\omega _n^2\gg T^2.$ In this limit
we find 
\begin{eqnarray}
\Pi _{\text{qu}}(q,\omega _n) &=&\frac T{\pi (q^2+\omega _n^2)}\ln \frac{qT}{%
q^2+\omega _n^2}  \label{Piqu} \\
&&+\frac 1{8\sqrt{q^2+\omega _n^2}},\,\,\,\,\,\,\,\,\,\, 
\begin{array}{c}
\,q^2+\omega _n^2\gg T^2
\end{array}
\nonumber
\end{eqnarray}

For $\omega _n=0$ and $q\ll T$ we obtain by analogy with the calculation of $%
\Pi (q,q_z,0)$ the result 
\begin{eqnarray}
I(q,q_z,0) &=&\frac T{4\pi }\frac 1{q^2+\alpha (1-\cos q_z)}  \label{I1} \\
&&\ \times \int\limits_0^1\frac{dz}{\sqrt{z(1-z)}}\frac{q^2z(1-z)+\alpha
z(1-\cos q_z)}{\left[ q^4z(1-z)+2\widetilde{\alpha }(q,q_z)\right] ^{3/2}}%
\,,\,\,\,\,\,\,\, 
\begin{array}{c}
q\ll T
\end{array}
\,\,\,\,\,\,\,  \nonumber
\end{eqnarray}
and its asymptotic form 
\begin{equation}
I(q,q_z,0)\simeq \frac T{2\pi q^4}\left[ \ln \frac{2q^2}\alpha -\frac{3+\cos
q_z}2\right] ,\,\,\,\,\,\,\alpha \ll \,q^2\ll T^2  \label{I1a}
\end{equation}
In the region with $q^2+\omega _m^2\gg \alpha $ we have 
\begin{eqnarray}
I(q,q_z,\omega _n) &=&\frac{Tq^2}{2\pi (q^2+\omega _m^2)^3}\ln \frac{%
2(q^2+\omega _m^2)^2}{q^2\alpha }  \label{I2} \\
&&\ +\frac T{4\pi }\frac{\omega _n^2-3q^2-(q^2+\omega _n^2)\cos q_z}{%
(q^2+\omega _n^2)^3}  \nonumber \\
&&\ +I_{\text{qu}}(q,\omega _n),\,\,\,\,\, 
\begin{array}{c}
q^2+\omega _n^2\gg \alpha
\end{array}
\nonumber
\end{eqnarray}
with 
\begin{eqnarray}
I_{\text{qu}}(q,\omega _n) &=&\frac T\pi \sum_{m\neq 0}\frac{q^2}{\left[
(\omega _n^2+q^2+2\omega _m\omega _n)^2+4q^2\omega _m^2\right] ^{3/2}} \\
&&\ \times \text{Arc}\tanh \frac{\omega _n^2+q^2+2\omega _m\omega _n}{\sqrt{%
(\omega _n^2+q^2+2\omega _m\omega _n)^2+4q^2\omega _m^2}}  \nonumber \\
&&\ +\frac T{4\pi }\sum_{m\neq 0}\frac{q^2+2\omega _m\omega _n+\omega _n^2}{%
\omega _m^2\left[ (\omega _n^2+q^2+2\omega _m\omega _n)^2+4q^2\omega
_m^2\right] }  \nonumber
\end{eqnarray}
In the ultraviolet limit $\,q^2+\omega _n^2\gg T^2$ the following
asymptotics takes place: 
\begin{eqnarray}
I_{\text{qu}}(q,\omega _n) &=&\frac{Tq^2}{\pi (q^2+\omega _n^2)^3}\ln \frac{%
qT}{q^2+\omega _n^2}  \label{Iqu} \\
&&\ +\frac T\pi \frac{\omega _n^2-q^2}{(q^2+\omega _n^2)^3}%
,\,\,\,\,\,\,\,\,\,\,\,\,\, 
\begin{array}{c}
\,\,\,q^2+\omega _n^2\gg T^2
\end{array}
\nonumber
\end{eqnarray}

\section*{Appendix B. Calculation of $1/N$-corrections to the constraint at $%
T\leq T_{\text{N\lowercase {eel}}}.$}

Consider briefly the calculation of the functions $R$ (\ref{R}) and $F$ (\ref
{F}) which determine, according to (\ref{Constr1/Nm}), the corrections to
the constraint to first order in $1/N$. First we introduce intermediate
cutoff parameters $C$ and $C^{\prime }$ determined by $\alpha ^{1/2}\ll C\ll
T\ll 2\pi C^{\prime }\ll \Lambda $ and divide the region of summation and
integration $q^2+\omega _n^2<\Lambda ^2$ into four regions:

\noindent 1. $\omega _n=0,$ $q<C.$

\noindent 2. $\omega _n=0,\,\,C<q<2\pi C^{\prime }$

\noindent 3. $\omega _n\neq 0,$ $q^2+\omega _n^2<2\pi C^{\prime }$

\noindent 4. $2\pi C^{\prime }<q^2+\omega _n^2<\Lambda $

\noindent Further we denote the contributions from $i$-th region to $R$ and $%
F$ as $R_i(T,x_{\overline{\sigma }})$ and $F_i(T,x_{\overline{\sigma }})$.

In the first region we can use the expressions for the functions $\Pi
(q,q_z,0)$ and $I(q,q_z,0)$ at $q^2\ll T$ (\ref{Pi1}), (\ref{I1}) and their
asymptotics (\ref{Pi1a1}), (\ref{Pi1a2}), (\ref{I1a}). Then 
\begin{eqnarray}
R_1(T,x_{\overline{\sigma }}) &=&\frac T{2\pi N}\ln \frac{2C^2}\alpha -\frac{%
(3+2x_{\overline{\sigma }})T}{4\pi N} \\
&&\ \times \ln \frac{\ln (2C^2/\alpha )+x_{\overline{\sigma }}}{x_{\overline{%
\sigma }}}+\frac T{4\pi N}I_1(x_{\overline{\sigma }})  \nonumber
\end{eqnarray}
where 
\begin{eqnarray}
I_1(x_{\overline{\sigma }}) &=&\frac 4N\int\limits_0^\infty
qdq\int\limits_{-\pi }^\pi \frac{dq_z}{2\pi }\left[ \frac 1{q^2+1-\cos q_z}%
\right. \\
&&\ \left. \times \frac{I(q,q_z,0)-\widetilde{\Pi }(q,q_z,0)}{\widetilde{\Pi 
}(q,q_z,0)}+\frac{3+2x_{\overline{\sigma }}}{2q^2}\frac{\theta (q^2-1/2)}{%
\ln (2q^2)+x_{\overline{\sigma }}}\right]  \nonumber
\end{eqnarray}
$\theta (x)\ $is the step function$.$ In the second and third regions we use
the expressions for the functions $\Pi (q,q_z,0)$ and $I(q,q_z,0)$ at $%
q^2\gg \alpha ,$ (\ref{Pi2}) and (\ref{I2}): 
\begin{eqnarray}
R_2(T,x_{\overline{\sigma }}) &=&\frac T{\pi N}\left[ \ln \frac TC-\frac{%
3+2x_{\overline{\sigma }}}4\ln \frac{4\pi \rho _s/NT}{\ln (2C^2/\alpha )+x_{%
\overline{\sigma }}}\right.  \nonumber \\
&&\ \left. +\ln \frac{2\pi C^{\prime }}T\frac{\ln (2T^2/\alpha )}{\ln
(2T^2/\alpha )+x_{\overline{\sigma }}}\right] \\
R_3(T,x_{\overline{\sigma }}) &=&\frac T{\pi N}\frac{\ln (2T^2/\alpha )}{\ln
(2T^2/\alpha )+x_{\overline{\sigma }}}\left[ \frac{4C^{\prime }}{3T}+\frac 12%
-\ln \frac{2\pi C^{\prime }}T\right]
\end{eqnarray}
where we have used the identity 
\begin{equation}
\ln (2T^2/\alpha )+x_{\overline{\sigma }}=4\pi \rho _s/NT  \label{id}
\end{equation}
which is satisfied in the zeroth order in $1/N$. In the fourth region we
obtain 
\begin{eqnarray}
R_4(T,x_{\overline{\sigma }}) &=&\frac{8T}{3\pi ^3N}\ln \frac{2T^2}\alpha
\ln \frac{N\Lambda }{16\rho _s}  \nonumber \\
&&\ -\frac{2T}{3\pi ^3N}\ln \frac{N\Lambda }{16\rho _s}-\frac{4C^{\prime }}{%
3\pi N}\frac{\ln (2T^2/\alpha )}{\ln (2T^2/\alpha )+x_{\overline{\sigma }}}
\end{eqnarray}
where we have used the asymptotic forms (\ref{Piqu}), (\ref{Iqu}) and the
identity (\ref{id}). After collecting all $R_i$ ($i=1...4$) the intermediate
cutoff parameters $C,C^{\prime }$ are canceled and we find the result (\ref
{R1}) of the main text.

Analogously, we obtain the contribution from the first region to $F$ in the
form 
\begin{equation}
F_1(T,x_{\overline{\sigma }})=\frac 1N\ln \frac{\ln (2C^2/\alpha )+x_{%
\overline{\sigma }}}{x_{\overline{\sigma }}}+I_2(x_{\overline{\sigma }})
\end{equation}
where 
\begin{eqnarray}
I_2(x_{\overline{\sigma }}) &=&\frac 4N\int\limits_0^\infty
qdq\int\limits_{-\pi }^\pi \frac{dq_z}{2\pi }\left[ \frac 1{(q^2+1-\cos
q_z)^2}\right.  \nonumber \\
&&\ \left. \times \frac 1{\widetilde{\Pi }(q,q_z,0)}-\frac 1{2q^2}\frac{%
\theta (q^2-1/2)}{\ln (2q^2)+x_{\overline{\sigma }}}\right]
\end{eqnarray}
Contributions from other three regions are also easily calculated: 
\begin{eqnarray}
F_2(T,x_{\overline{\sigma }}) &=&\frac 1N\ln \frac{4\pi \rho _s/NT}{\ln
(2C^2/\alpha )+x_{\overline{\sigma }}} \\
F_3(T,x_{\overline{\sigma }}) &=&{\cal O(}1/\ln (2T_{\text{Neel}}^2/\alpha ))
\\
F_4(T,x_{\overline{\sigma }}) &=&\frac 8{\pi ^2N}\ln \frac{N\Lambda }{16\rho
_s}
\end{eqnarray}
Summing up all $F_i$ ($i=1...4$) we find the result (\ref{F1}) of the main
text.

\section*{Appendix C. The order parameter and transition temperature at $
\lowercase {d=2+\varepsilon } $}

In this Appendix we consider the calculation of the sublattice magnetization
to first order in $1/N$ in the space with the dimensionality $%
d=2+\varepsilon .$ We will be interested in the terms of the leading order
in $\varepsilon $ at not too small temperatures $T\gg Je^{-1/\varepsilon }$
(which is an analog of the renormalized classical regime in the 2D case), so
that only the contributions with zero Matsubara frequences will be taken
into account. Consider first the results for the functions $\Pi $ and $I.$
Evaluating the integrals in (\ref{Pi}) and (\ref{I}) at an arbitrary space
dimensionality $2<d<4$ (see, e.g., Ref. \cite{Ryder} for the procedure of
calculation of such integrals) we have 
\begin{eqnarray}
\Pi (q,0) &=&\frac{TK_dA_d}{q^{4-d}} \\
I(q,0) &=&\frac{TK_dA_d(3-d)}{q^{6-d}}
\end{eqnarray}
where $q$ is the $d$-dimensional vector, 
\begin{eqnarray}
A_d &=&\frac{\Gamma (d/2)\Gamma (2-d/2)\Gamma ^2(d/2-1)}{2\Gamma (d-2)} \\
K_d^{-1} &=&2^{d-1}\pi ^{d/2}\Gamma (d/2)  \nonumber
\end{eqnarray}
$\Gamma (x)$ is the Euler gamma-function. At $d=2+\varepsilon $ we find to
leading order in $\varepsilon $ 
\begin{equation}
\Pi (q,0)=q^2I(q,0)=\frac{2TK_2}{\varepsilon q^{2-\varepsilon }},\,\,\,K_2=%
\frac 1{2\pi }
\end{equation}
The constraint equation to first order in $1/N$ (\ref{Constr1/Nm}) takes the
form 
\begin{eqnarray}
&&\ \ 1-\frac{gT^{1+\varepsilon }K_2}\varepsilon (1-\frac 2N)  \nonumber \\
\ &=&\overline{\sigma }^2\left[ 1-\frac{K_2}N\ln \frac{T^{1+\varepsilon
}/\varepsilon +\overline{\sigma }^2/g}{\overline{\sigma }^2/g}\right]
\label{c2+e}
\end{eqnarray}
Using the identity 
\begin{equation}
gT^{1+\varepsilon }/\varepsilon +\overline{\sigma }^2=1
\end{equation}
which is satisfied in the zeroth order in $1/N$ and transforming the
logarithmic term in (\ref{c2+e}) into a power, we obtain 
\begin{eqnarray}
\overline{\sigma } &=&\left( 1-T^{1+\varepsilon }/T_{\text{Neel}%
}^{1+\varepsilon }\right) ^{\beta _{2+\varepsilon }}  \label{m2+e} \\
\beta _{2+\varepsilon } &=&(1+1/N)/2+{\cal O}(1/N^2,\varepsilon )
\label{b2+e}
\end{eqnarray}
The Neel temperature is determined by 
\begin{equation}
T_{\text{Neel}}^{1+\varepsilon }=\frac \varepsilon {gK_2(1-2/N)}=\frac{2\pi
\rho _s\varepsilon }{N-2}  \label{tne}
\end{equation}
This result coincides with the result of the RG analysis \cite{Brezin}. The
RG\ result for the critical exponent $\beta $ reads 
\begin{equation}
\beta _{2+\varepsilon }=\frac 12\left( 1+\frac 1{N-2}\right) +{\cal O}%
(\varepsilon )
\end{equation}
Thus one have to replace $N\rightarrow N-2$ in\ (\ref{b2+e}). Such a
replacement is analogous to this in the renormalized classical regime of
Ref. \cite{Chubukov} and may be justified by the calculations of terms of
order of $1/N^2$, which we did not carry out. As demonstrated in Ref. \cite
{Chubukov} by calculations of analogous contributions up to $1/N^2$, this
replacement should be indeed performed. Since the denominator in (\ref{tne})
is of order of $N,$ the replacement $N\rightarrow N-2$ occurs already in the
first-order expression for the transition temperature (\ref{tne}).

According to (\ref{m2+e}), two regimes are possible in the temperature
dependence of the order parameter. At $Je^{-1/\varepsilon }\ll T\ll T_{\text{%
Neel}}$ we have the spin-wave behavior 
\begin{equation}
\overline{\sigma }=1-\frac{(N-1)T^{1+\varepsilon }}{4\pi \rho _s\varepsilon }
\end{equation}
For $N=3$ this result is analogous to the quasi-2D case result (\ref
{MagnSSWT}) in the quantum spin case. At $1-T/T_{\text{Neel}}\ll 1$ the
temperature dependence of the sublattice magnetization changes from the
linear one to the power behavior with the critical exponent $\beta
_{2+\varepsilon }.$

Two above temperature regimes correspond to different pictures of the
excitation spectrum. In the low-temperature regime $T\ll T_{\text{Neel}}$ we
have from (\ref{GL}) at quasimomenta $q<T$ (only such $q$ give a
contribution to thermodynamic quantities) the zeroth order longitudinal
Green's function 
\begin{equation}
G_l^{N=\infty }(q,0)=\frac g2\Pi (q,0)=\frac{gTK_2}{\varepsilon
q^{2-\varepsilon }}
\end{equation}
which corresponds to spin-wave excitations. Near the phase transition point
we have at an arbitrary $q$ (except for the exponentially-narrow
hydrodynamic region $q<(2\overline{\sigma }^2/g)^{1/\varepsilon }$) 
\begin{equation}
G_l^{N=\infty }(q,0)=\frac 1{q^2}
\end{equation}
which corresponds to critical (non-spin-wave) excitations.

{\sc Figure caption}

Fig.1. The theoretical temperature dependences of the relative sublattice
magnetization $\overline{S}/\overline{S}_0$ from different spin-wave
approximations and from the $1/N$-expansion in the $O(N)$ model (equations (%
\ref{Constr2D}) and (\ref{MagnCr})), and the experimental points for La$_2$%
CuO$_4$ (Ref.\cite{Keimer}).

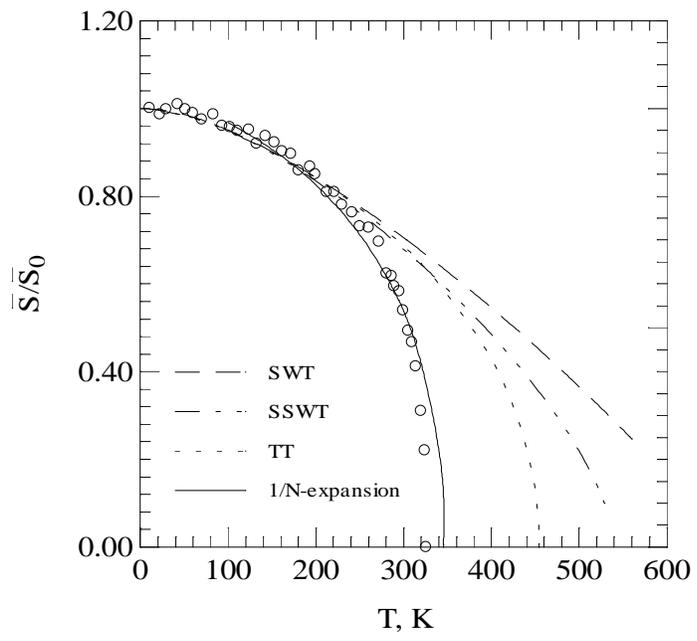
\end{document}